\documentclass[aps,floats,amssymb,amsmath,twocolumn]{revtex4}
\usepackage{multirow}
\usepackage{calc}
\usepackage{psfrag}
\usepackage{graphicx}
\usepackage{color}

\begin{document}

\begin{abstract}
The Bloch theorem enables reduction of the eigenvalue problem of the single-particle Hamiltonian that commutes with translational group. Based on a group theory analysis we present generalization of the Bloch theorem that incorporates all additional symmetries of a crystal. The generalized Bloch theorem constrains the form of the Hamiltonian which becomes manifestly invariant under additional symmetries.
In the case of isotropic interactions the generalized Bloch theorem gives a unique Hamiltonian. This Hamiltonian coincides with the Hamiltonian in the periodic gauge.
In the case of anisotropic interactions the generalized Bloch theorem allows a family of Hamiltonians. Due to the continuity argument we expect that even in this case the Hamiltonian in the periodic gauge defines observables, such as Berry curvature, in the inverse space. For both cases we present examples and demonstrate that the average of the Berry curvatures of all possible Hamiltonians in the Bloch gauge is the Berry curvature in the periodic gauge.
\end{abstract}
\title{Generalized Bloch theorem and topological characterization}
\author{E. Dobard{\v{z}}i\'c$^1$}
\author{M. Dimitrijevi\'c$^1$}
\author{M. V. Milovanovi\'c$^2$}
\affiliation{$^1$ Faculty of Physics, University of Belgrade, 11001 Belgrade, Serbia\\
$^2$ Scientific Computing Laboratory, Institute of Physics
Belgrade, University of Belgrade, Pregrevica 118, 11 080 Belgrade,
Serbia}
\maketitle

\section{Introduction}
The Bloch theorem~\cite{Bloch-Physik.52.555} is very useful in description of a quasi-particle (electron, phonon, etc.) in crystals. The solution of the Bloch problem can be stated in a representation-independent form. Namely, the Bloch vector $|\psi(\mathbf{k})\rangle$ can be written as a linear superposition of the localized, in the unit cell $\mathbf{R}$ at position $\mathbf{a}_j$, orbitals $|\mathbf{R},\mathbf{a}_j\rangle$ of a crystal with coefficients, $c_j(\mathbf{k})$, that depend on Bloch momentum $\mathbf{k}$ and type of orbital $j$ in the unit cell:
\begin{equation}\label{EBlochLin}
|\psi(\mathbf{k})\rangle = \sum_{\mathbf{R},j} c_j(\mathbf{k}) \mathrm{e}^{\mathrm{i}\mathbf{k}\cdot\mathbf{R}} |\mathbf{R},\mathbf{a}_j\rangle.
\end{equation}
It is obvious that this form satisfies the Bloch theorem
$T_{\mathbf{R}'} |\psi(\mathbf{k})\rangle = \exp(-\mathrm{i}\mathbf{k}\cdot\mathbf{R}') |\psi(\mathbf{k})\rangle$, where $T_{\mathbf{R}'}$ is the translation operator for the vector $\mathbf{R}'.$

In this work we present a generalization of the Bloch theorem that incorporates all additional (beside translational) symmetries of a crystal. The generalized Bloch theorem constrains the form of the Hamiltonian which becomes manifestly invariant under additional symmetries.

Any non-Bravais lattice Hamiltonian is invariably multidimensional (always matrix), and may be affected by unitary (matrix) transformations, i.e. may transform into various ``gauges". Different choices may lead to the same topological invariants (Chern number, second Chern number, and the gauge concept is very useful in characterizing topology) but when we discuss some physical observables that concern ``geometry" i.e. the exact configuration of the lattice it is natural to work with a special Hamiltonian and thus use so called ``periodic gauge". We show, by using group theory techniques, that this gauge is constrained by symmetry arguments.

In the following we would like to illustrate the importance of choosing the right gauge. The coefficients $c_j(\mathbf{k})$ in Eq.~(\ref{EBlochLin}) naturally represent the Bloch vector in the space of Bloch momentum, and therefore can be used in the definition of Berry connection. Berry connection provides a way to describe how the phase of the Bloch vector varies as we change Bloch momentum. The choice of the coefficients in the expansion of $|\psi(\mathbf{k})\rangle$ is by no means unique - they are fixed, up to overall $U(1)$ phase ($j$ independent but possible $\mathbf{k}$ dependent), by the form of the Hamiltonian we work with. There is a freedom of choosing the phases of coefficients connected by non-trivial diagonal unitary transformations, but two choices - two forms of Hamiltonians stand out. The choice in Eq.~(\ref{EBlochLin}) implies Berry connection
$$
\mathbf{A}^B = \mathrm{i} \sum_j c_j^*(\mathbf{k}) \nabla_{\mathbf{k}} c_j(\mathbf{k}),
$$
where $B$ stands for ``Bloch gauge". A different choice,
$$
|\psi(\mathbf{k})\rangle = \sum_{\mathbf{R},j} u_j(\mathbf{k}) \mathrm{e}^{\mathrm{i}\mathbf{k}\cdot(\mathbf{R}+\mathbf{a}_j)} |\mathbf{R},\mathbf{a}_j\rangle,
$$
on the other hand, implies
$$
\mathbf{A}^P = \mathrm{i} \sum_j u_j^*(\mathbf{k}) \nabla_{\mathbf{k}} u_j(\mathbf{k}),
$$
where $P$ stands for ``periodic gauge".
These are the most natural choices; they coincide with two possibilities to define the Fourier transform in the space of the Bloch vectors:
$
|\mathbf{k},j\rangle_B = \sum_{\mathbf{R}} \mathrm{e}^{\mathrm{i}\mathbf{k}\cdot\mathbf{R}} |\mathbf{R},\mathbf{a}_j\rangle$
and
$|\mathbf{k},j\rangle_P = \sum_{\mathbf{R}} \mathrm{e}^{\mathrm{i}\mathbf{k}\cdot(\mathbf{R}+\mathbf{a}_j)} |\mathbf{R},\mathbf{a}_j\rangle
$~\cite{NJPhys.11.095003}, and thus two Hamiltonians in inverse space.
Although physics should be invariant under these choices, and we speak of two gauges, we will show in examples that they define two different Berry curvatures. So the word gauge in this sense is a misnomer and we must look for a ``physical gauge". In this gauge we would be certain that we are defining a physical Berry curvature, a suitable candidate for the intrinsic magnetic field in the inverse space. This would define a local magnetic length that is important to know for example in the case of fractional physics in the bands with non-zero Chern number. There are proposals~\cite{arXiv.1407.5635,Dmitry-PhysRevLett.110.165304} how to measure Berry curvature in experiments, and thus it is important to know in which gauge it is defined.

From the Karplus-Luttinger argument~\cite{PhysRev.95.1154,RevModPhys.82.1959} that the coordinate operator in the inverse space can be represented as
$$
\mathbf{r} \rightarrow -\mathrm{i}\nabla_{\mathbf{p}} + A^P
$$
we expect that the periodic gauge is the right choice. This choice in the literature is commonly used without a complete understanding why it is physical. On the other hand, there are references that use the Bloch gauge (and reach wrong expectations). By establishing the
generalized Bloch theorem we provide strong arguments that the periodic gauge is the natural choice given that it, i.e. the Hamiltonian that this gauge defines, respects the symmetries of the underlying lattice.

Thus the symmetry is a determining factor that provides the right choice. In the past researchers were finding in certain examples that the Berry curvature calculated in the periodic gauge is the most symmetric~\cite{EurPhysJ.77.351,EPL.106.60002,PhysRevB.89.235424}. Here we explore the symmetry aspect by using the techniques of the group theory, and finding the form of the Hamiltonian in which any symmetry action can be made explicit. In most cases, after simple
rewritings, the manifestly invariant Hamiltonian is equal to the Hamiltonian in the periodic gauge.

We demonstrate that in the case of the isotropic interactions (which will be precisely defined below) the Hamiltonian that respects the symmetries of the lattice is uniquely defined by the generalized Bloch theorem and coincides with the Hamiltonian in the periodic gauge. In the case of anisotropic interactions we find, due to a phase freedom in the construction, a family of generalized Bloch Hamiltonians that respect the symmetries of the lattice. One of the Hamiltonians corresponds to the periodic gauge choice. This choice seems a natural choice even in this case of anisotropic interactions because it can be viewed as an extension of the isotropic case. In addition we find that the average Berry curvature of the eigenstates of all possible Bloch Hamiltonians is equal to the Berry curvature in the periodic gauge.

The paper is organized as follows: In Sec. II we review the Bloch theorem and its Hamiltonian
construction. In Sec. III, to begin the generalization, we assume an existence of a basis in which the action of crystal symmetries is made explicit, i.e. a symmetry adapted basis. In the same section we seek for a realization - a basis defined by a special class of projectors that represent an invariance of the action of symmetries, and reach the form of the Hamiltonian in that basis. In Sec. IV, we find explicit expressions for the vectors of the basis and therefore Hamiltonian. In the same section we discuss when the Hamiltonian that respects all additional symmetries beside translation is unique or represents a family of Hamiltonians. In Sec. V, to illustrate the formal procedure described in the previous sections (III and IV), we discuss concrete examples of lattices and constructions of the generalized Bloch Hamiltonians. Sec. VI is a short summary.

\section{Bloch theorem}
A crystal is the system of periodically arranged atoms along $d$ linearly independent directions. A part of a crystal, called unit cell (UC), contains minimal set of atoms sufficient to build the whole crystal by the action of discrete translations for the Bravais lattice vectors $\mathbf{R}=\sum_{p=1}^d n_p\mathbf{b}_p$, where $n_p$ are integers and $\mathbf{b}_p$ are called primitive vectors of direct lattice. Quasi-particle basis
$
|\mathbf{k},j\rangle
$
($j$ enumerates basis vectors - orbitals, $j=1,\dots,J$) is described by the quasi-momentum quantum numbers $\mathbf{k}=\sum_{p=1}^d k_p\mathbf{g}_p$, where $k_p\in(-1/2,1/2]$ and $\mathbf{g}_p$ are called primitive vectors of reciprocal lattice ($\mathbf{b}_p\cdot\mathbf{g}_q=2\pi\delta_{pq}$, $\delta_{pq}$ is Kronecker delta). The Bloch theorem states that the basis vector
$
|\mathbf{k},j\rangle
$
translated for the Bravais vector $\mathbf{R}$ changes as
\begin{equation}\label{EBlochpsi}
T_{\mathbf{R}} |\mathbf{k},j\rangle = \mathrm{e}^{-\mathrm{i}\mathbf{k}\cdot\mathbf{R}} |\mathbf{k},j\rangle
\end{equation}
where $T$ is the translation operator. In other words, if the $j$th basis vector in the zeroth unit cell ($\mathbf{R} = \mathbf{0}$) is $|\mathbf{0},\mathbf{a}_j\rangle$, then in the unit cell $\mathbf{R}$ it is $\mathrm{e}^{\mathrm{i}\mathbf{k}\cdot\mathbf{R}} |\mathbf{R},\mathbf{a}_j\rangle$ and consequently $|\mathbf{k},j\rangle = \sum_{\mathbf{R}} \mathrm{e}^{\mathrm{i}\mathbf{k}\cdot\mathbf{R}} |\mathbf{R},\mathbf{a}_j\rangle$. Vectors $|\mathbf{0},\mathbf{a}_j\rangle$ and $|\mathbf{R},\mathbf{a}_j\rangle$ are identical, but belong to different blocks, $\mathbf{0}$ and $\mathbf{R}$, of the vector $|\mathbf{k},j\rangle$, respectively.

Reduction of operators describing quasi-particle in a crystal such as the Hamiltonian, dynamical matrix, overlap matrix, etc. invariant under discrete translations, is straightforward with the help of Eq.~(\ref{EBlochpsi}).
Any Hamiltonian eigenvector $|\psi(\mathbf{k})\rangle$ can be expressed as a linear combination with coefficients [Eq.~(\ref{EBlochLin})] $c_{j}(\mathbf{k})$
\begin{equation}\label{EFunBloch}
|\psi(\mathbf{k})\rangle = \sum_{j} c_{j}(\mathbf{k}) |\mathbf{k},j\rangle = \sum_{\mathbf{R},j} c_{j}(\mathbf{k}) \mathrm{e}^{\mathrm{i}\mathbf{k}\cdot\mathbf{R}} |\mathbf{R},\mathbf{a}_j\rangle.
\end{equation}
Now, the eigenvalue problem of the Hamiltonian can be replaced by the set of homogeneous equations in coefficients $c_{j}(\mathbf{k})$,
\begin{equation}\label{EHomBloch}
\sum_{j'} \left[ \sum_{\mathbf{R}'} \mathrm{e}^{\mathrm{i}\mathbf{k}\cdot\mathbf{R}'} H_{\mathbf{R}',j'}^{\mathbf{0},j} -  E(\mathbf{k})\delta_{j,j'}\right] c_{j'}(\mathbf{k}) = 0,
\end{equation}
for all $j,$ where $H_{\mathbf{R}',j'}^{\mathbf{0},j}$ are Hamiltonian matrix elements in the basis $|\mathbf{R},\mathbf{a}_j\rangle$,
\begin{equation}\label{EBlochHam}
H_{\mathbf{R}',j'}^{\mathbf{0},j} = \langle\mathbf{0},\mathbf{a}_j| H |\mathbf{R}',\mathbf{a}_{j'}\rangle.
\end{equation}
This set of equations can be written in the form of the eigenvalue problem of the Bloch Hamiltonian $H(\mathbf{k})$
\begin{equation}\label{EBlochRedHam}
H(\mathbf{k}) c(\mathbf{k}) = E(\mathbf{k}) c(\mathbf{k}),\quad
H(\mathbf{k}) = \sum_{\mathbf{R}} \mathrm{e}^{\mathrm{i}\mathbf{k}\cdot\mathbf{R}} H^{\mathbf{0}}_{\mathbf{R}},
\end{equation}
where $c(\mathbf{k})$ is a column with components $c_{j}(\mathbf{k})$ and $H_{\mathbf{R}}^{\mathbf{0}}$ is matrix having matrix elements $H_{\mathbf{R},j'}^{\mathbf{0},j}(\mathbf{k})$. In some approximations, e.g. tight-binding (not using L\"owdin~\cite{Lowdin-JChemPhys.18.365} nor Wannier~\cite{Wannier-PhysRev.52.191,Wannier-RevModPhys.34.645} functions), the Kronecker delta in Eq.~(\ref{EHomBloch}) is replaced by the overlap integrals, $S_{\mathbf{R}',j'}^{\mathbf{0},j} = \langle\mathbf{0},\mathbf{a}_j|\mathbf{R}',\mathbf{a}_{j'}\rangle$, and the Bloch reduced overlap matrix $S(\mathbf{k}) = \sum_{\mathbf{R}} \mathrm{e}^{\mathrm{i}\mathbf{k}\cdot\mathbf{R}} S_{\mathbf{R}}^{\mathbf{0}}$ appears on the right hand side of Eq.~(\ref{EBlochRedHam}).

There are many single-particle approximations~\cite{Ashcroft1976Symmetry, Kittel1986Introduction, Tinkham1964Group, Dresselhaus2007Group} which can be solved by using the Bloch theorem. In the nearly free electron approximation the basis consists of functions $\langle\mathbf{r}|\mathbf{k},\mathbf{G}\rangle = \psi^{(\mathbf{k})}_{\mathbf{G}}(\mathbf{r}) = \mathrm{e}^{\mathrm{i}(\mathbf{k}-\mathbf{G})\cdot\mathbf{r}}$ ($\mathbf{G}=\sum_{i=1}^d m_i\mathbf{g}_i$ are vectors of reciprocal lattice and $m_i$ are integers), and the finite set of reciprocal lattice vectors $\mathbf{G}$ depends on the physical situation. The tight binding approximation deals with basis $|\mathbf{k},\alpha nlm\rangle$, which are electron orbitals with quantum numbers $nlm$ on the atom $\alpha$. In this approximation, previously used symbol $j$ in enumerating basis is replaced by the four numbers $(\alpha nlm)$. In the harmonic approximation basis $|\mathbf{k},\alpha i\rangle$ are orthogonal unit vectors $\mathbf{e}_i$ ($i=x,y,z$) on atom $\alpha$.

\section{Generalization}
To begin with generalization, let $\mathcal{G}$ be the symmetry group of a crystal, and let $\Gamma$ (corresponding to the quasi-momentum quantum number $\mathbf{k}$ in the Bloch theorem) label unitary irreducible representation (IR) $D^{(\Gamma)}(A)$. The basis for IR $\Gamma$ is given by the vectors $|\Gamma,j,a\rangle$ which fulfill
\begin{equation}\label{ESAB}
D(A) |\Gamma,j,a\rangle = \sum_{b=1}^{d_{\Gamma}} D^{(\Gamma)}_{ba}(A) |\Gamma,j,b\rangle.
\end{equation}
Here, $D^{(\Gamma)}_{ba}(A)$ is matrix element ($b,a$) of the $d_{\Gamma}$-dimensional irreducible representation $D^{(\Gamma)}(A)$ and $D(A)$ is unitary representation of the group element $A$. The definition (\ref{ESAB}) is viewed as a starting point by some authors and as a consequence by others~\cite{Wigner1959Group,
Landau1977Quantum, Elliot1979Symmetry,
Tinkham1964Group, Dresselhaus2007Group}. The basis~(\ref{ESAB}) is called symmetry adapted basis (SAB). If the group $\mathcal{G}$ consists of translations only, IRs are one dimensional, $D^{(\mathbf{k})}(\mathbf{R}) = \mathrm{e}^{-\mathrm{i}\mathbf{k}\cdot\mathbf{R}}$. This is the Bloch theorem, Eq.~(\ref{EBlochpsi}).
Vectors differing only in $a$ are called partners. It is also said that vectors $|\Gamma,j,a\rangle$ for fixed $\Gamma$ and $j$ belong to one multiplet.

In the previous paragraph we introduced usual approach of the group theory with the basis vector, $|\Gamma,j,a\rangle$, that we will work with, in order to incorporate symmetries beside translation in the quantum-mechanical description of the problem defined on the lattice. The introduction of $|\Gamma,j,a\rangle$ corresponds to the introduction of $|\mathbf{k},j\rangle$ in the case of ordinary Bloch theorem, see Eq.~(\ref{EBlochpsi}). The extra label $a$ enables us to track the action of the orthogonal transformations that leave an origin unchanged. Namely, by inspecting the symmetry of the lattice we choose a unit cell and a point of the highest symmetry i.e. the origin. (For example, in the case of the hexagonal lattice, we may choose for such a point any of two atoms (A or B) in a unit cell. In general an origin may not be chosen on an atom.) Once we choose the origin we may inspect how the vectors of the atoms in the unit cell to which the origin belongs, $\mathbf{a}_j; j=1,\dots,J$, transform under orthogonal transformations. Then we inspect the sets of all the atoms connected to the basis vectors $\mathbf{a}_j$ (which may not belong to the same unit cell and are connected by the orthogonal transformations). The number of elements of the set with the largest number of elements, defines $d_{\Gamma}$-the dimension of the irreducible representation $\Gamma$. (In the case of the hexagonal lattice $d_{\Gamma}=3$ related to three neighboring (to the atom at the origin) atoms that transform into each other by $C_3$ symmetry.)

Similarly to the case of the Bloch theorem where we have the decomposition of the basis vector $|\mathbf{k},j\rangle$ over site orbitals $|\mathbf{R},\mathbf{a}_j\rangle$, $|\mathbf{k},j\rangle = \sum_{\mathbf{R}}\exp(\mathrm{i}\mathbf{k}\cdot\mathbf{R}) |\mathbf{R},\mathbf{a}_j\rangle$, we have a decomposition of basis vectors $|\Gamma,j,a\rangle$ in the case of the generalized Bloch theorem. First we note that we can write the Bloch basis vector as $|\mathbf{k},j\rangle = \sum_{\mathbf{R}'}\exp(\mathrm{i}\mathbf{k}\cdot\mathbf{R}') |\mathbf{R}'+\mathbf{R},\mathbf{a}_j\rangle$, which simply means that instead of the reference point $\mathbf{0}$ we choose $\mathbf{R}$ i.e. an arbitrary fixed unit cell vector. To understand the decomposition for $|\Gamma,j,a\rangle$ we introduce $|\mathbf{R}'+\mathbf{R},\mathbf{a}_j,a\rangle = \exp(\mathrm{i}\mathbf{k}\cdot\mathbf{R}') |\mathbf{R}'+\mathbf{R},\mathbf{a}_j\rangle$ so that we can rewrite $|\mathbf{k},j\rangle$ in the following form
\begin{equation}\label{EBlochForm}
|\mathbf{k},j,a\rangle = \sum_{\mathbf{R}'} |\mathbf{R}'+\mathbf{R},\mathbf{a}_j,a\rangle.
\end{equation}
Here $a=1$ (the index is obsolete in this case) and the sum denotes the sum over all translations i.e. group elements. This rewriting shows in a transparent way the physical view of the Bloch theorem: by knowing the orbital at arbitrary fixed $\mathbf{R}$ we can easily reconstruct the other orbitals by the action of translational symmetry. Now, similarly to the Bloch case we can represent $|\Gamma,j,a\rangle$ as
\begin{equation}\label{EGenBlochForm}
|\Gamma,j,a\rangle = \sum_{X} |X\mathbf{R},X\mathbf{a}_j,a\rangle,
\end{equation}
where the sum is over group elements $X$ and $(\mathbf{R},\mathbf{a}_j)$ is an arbitrary site position and plug Eq.~(\ref{EGenBlochForm}) in Eq.~(\ref{ESAB}).

Note that Eq.~(\ref{EGenBlochForm}) is a generalization of the Bloch case in which with one atom (site) we associate a set of orbitals defined by the ways i.e. symmetry operations that we use to reach the atom in question from the origin $\mathbf{R}$. So the ket $|X\mathbf{R},X\mathbf{a}_j,a\rangle$ cannot be identified by a position (only) but  as a vector-orbital defined also by symmetry operation $X$. Thus the sum in Eq.~(\ref{EGenBlochForm}) should not be viewed as an ordinary sum but as a direct sum of vectors. In this way we can view vector $|\Gamma,j,a\rangle$ as a column of vectors $|X\mathbf{R},X\mathbf{a}_j,a\rangle$ and reach conclusions below.

Besides permutational representations in $D$  on the left-hand side of Eq.~(\ref{ESAB}), there are representations $D_{j}$ that act only on a general ket $|X\mathbf{R},X\mathbf{a}_j,a\rangle$ in $|\Gamma,j,a\rangle$. In this case a group element may change also zeroth unit cell, i.e. $|A\mathbf{0},\mathbf{a}_j,a\rangle \neq |\mathbf{0},\mathbf{a}_j,a\rangle$. Representation $D_{j}$ may be the representation $D^{(l,(-1)^l)}$ for the atomic orbital of electron with angular momentum $l$, the vector representation for phonons, etc. We decompose $D(A)$ into permutational part, $D^{P}(A)$, and $D_j(A)$ i.e. $D(A)=D^P(A)D_j(A)$. Acting by $D^P(A^{-1})$ from the left-hand side we have
$$
D_{j}(A) |X\mathbf{R},X\mathbf{a}_j,a\rangle = \sum_{b=1}^{d_{\Gamma}} D^{(\Gamma)}_{ba}(A) |AX\mathbf{R},AX\mathbf{a}_j,b\rangle,
$$
for any group element $X$ which we may choose to be identity. Thus
$$
D_{j}(A) |\mathbf{R},\mathbf{a}_j,a\rangle = \sum_{b=1}^{d_{\Gamma}} D^{(\Gamma)}_{ba}(A) |A\mathbf{R},A\mathbf{a}_j,b\rangle.
$$
Here the group element $A$ transfers the atom $\mathbf{a}_j$ from the unit cell $\mathbf{R}$, i.e. $(\mathbf{R},\mathbf{a}_j)$, to the atom $(A\mathbf{R},A\mathbf{a}_j)$. Depending on the group element $A$, it could happen that $A\mathbf{R}$ and $\mathbf{R}$ are the same unit cells and/or $A\mathbf{a}_j$ and $\mathbf{a}_j$ are the same orbitals. The previous transformation rule can be rewritten (remembering that all the representations are unitary) as the generalization of the Bloch theorem
\begin{subequations}\label{EgBt}
\begin{equation}\label{EgBtT}
|A\mathbf{R},A\mathbf{a}_j,a\rangle = \sum_{b=1}^{d_{\Gamma}} {D^{(\Gamma)}_{ab}}^*(A) D_{j}(A) |\mathbf{R},\mathbf{a}_j,b\rangle,
\end{equation}
where the asterisk stands for the complex conjugation. This effectively means that to each site in a unit cell $\mathbf{R}$ and position specified by the basis vector $\mathbf{a}_j$ we associate (ordered multiple i.e. $d_{\Gamma}$-tuple) $\{|\mathbf{R},\mathbf{a}_j,a\rangle;a=1,\dots,d_{\Gamma}\}$ (which under orthogonal transformations $A$ is transformed by the matrices of the irreducible representation $\Gamma$-$D^{(\Gamma)}_{ab}$ in Eq.~(\ref{EgBt})). Notice that for crystals with only the translational symmetry group, the representation $D_{j}(\mathbf{R})$ is always the identity operator and $A\mathbf{a}_j = \mathbf{a}_j$. This gives the Bloch theorem, Eq.~(\ref{EBlochpsi}), i.e. $|\mathbf{R}+\mathbf{R}',\mathbf{a}_j\rangle = \mathbf{e}^{\mathrm{i}\mathbf{k}\cdot\mathbf{R}'} |\mathbf{R},\mathbf{a}_j\rangle$.

In the Bloch theorem, it is sufficient to define quasi-particle orbitals and construct the Bloch Hamiltonian. In our generalization of the Bloch theorem only certain, restricted by symmetry, linear combinations of orbitals are allowed. To construct these linear combinations, it is sufficient to extract group elements $A$ that do not transfer the orbital $(\mathbf{R},\mathbf{a}_j)$, i.e. $(A\mathbf{R},A\mathbf{a}_j) = (\mathbf{R},\mathbf{a}_j).$
These elements form the subgroup $\mathcal{S}_{\mathbf{R},j}$ of the group $\mathcal{G}$ called stabilizer or little group of the orbital $(\mathbf{R},\mathbf{a}_j)$. In other words, atom $\mathbf{a}_j$ in the unit cell $\mathbf{R}$ is the fixed point of the action of its stabilizer, $\mathcal{S}_{\mathbf{R},j}$. The dimension (the number of the group elements of $\mathcal{S}_{\mathbf{R},j}$) we label by $s_j$.

After introducing stabilizers, we are able to find the allowed linear combinations of orbitals. Let us build the column $|\mathbf{R},\mathbf{a}_j\rangle\rangle$ of the vectors $|\mathbf{R},\mathbf{a}_j,a\rangle$, i.e. $d_{\Gamma}$-tuples. We use this special ket notation, $|\dots\rangle\rangle$, to differentiate the introduced column vector from ordinary vector - orbital, $|\mathbf{R},\mathbf{a}_j\rangle$, usually used. Then, Eq.~(\ref{EgBtT}) can be written in the form $|A\mathbf{R},A\mathbf{a}_j\rangle\rangle = \left[{D^{(\Gamma)}}^*(A) \otimes D_{j}(A)\right] |\mathbf{R},\mathbf{a}_j\rangle\rangle$, where ``$\otimes$" stands for tensor product operator. Summing over the stabilizer elements $A\in \mathcal{S}_{\mathbf{R},j}$ of the previous equality we obtain projectors $P^{(\Gamma)}_{\mathbf{R},j}$,
\begin{align}\label{EgBtS}
&P^{(\Gamma)}_{\mathbf{R},j} |\mathbf{R},\mathbf{a}_j\rangle\rangle = |\mathbf{R},\mathbf{a}_j\rangle\rangle,\nonumber\\
&P^{(\Gamma)}_{\mathbf{R},j} = \frac{1}{s_{j}} \sum_{A\in \mathcal{S}_{\mathbf{R},j}}{D^{(\Gamma)}}^*(A) \otimes D_{j}(A).
\end{align}\end{subequations}
These projectors are the identity operators in the Bloch theorem, since the stabilizer there consists of only the identity group element. Therefore the projectors $P^{(\Gamma)}_{\mathbf{R},j}$ are unnecessary therein. In our generalization, the projectors~(\ref{EgBtS}) restrict the linear combinations of the basis $|\mathbf{R},\mathbf{a}_j,a\rangle$ $a=1,\dots,d_{\Gamma}$ to their range. This basis is choosen to have orthogonal columns with the non-zero components $\sqrt{s_j}|\mathbf{R},\mathbf{a}_j,a\rangle$ and unit length of ket. If the projector's range is multidimensional ($\mathrm{Tr}P^{(\Gamma)}_{\mathbf{R},j} > 1$), it can be written, re-enumerating components if necessary, as a direct sum of one-dimensional projectors. Then the vectors are taken from the range of these one-dimensional projectors complemented by zeroes in the rest of the space, ensuring in this way vectors' orthogonality.

From the Eq.~(\ref{EgBtT}) we see that the group element $A\notin \mathcal{S}_{\mathbf{R},j}$, transfers the vectors $|\mathbf{R},\mathbf{a}_j\rangle\rangle$ to the other vector $|A\mathbf{R},A\mathbf{a}_j\rangle\rangle\neq|\mathbf{R},\mathbf{a}_j\rangle\rangle$ making it unnecessary to look for other reduced projectors. A sufficient set of reduced projectors is for the atoms not connected by the group action, i.e. for $(\mathbf{R},\mathbf{a}_j)$ and $(\mathbf{R}',\mathbf{a}_{j'})$ if $(\mathbf{R},\mathbf{a}_j) \neq ({A\mathbf{R}'},A\mathbf{a}_{j'})$ for all $A\in \mathcal{G}$. Such atoms are called the orbit representatives of the crystal for the symmetry group $\mathcal{G}$ and are not uniquely defined. However, this has no influence on the results. Although this generalization holds for any choice of orbit representatives, even within different unit cells, we will take them from the same unit cell, $\mathbf{R}' = \mathbf{R} = \mathbf{0}$. Hereafter, index $\mathbf{0}$ in $P^{(\Gamma)}_{\mathbf{0},j}$ and $\mathcal{S}_{\mathbf{0},j}$ will be omitted.

We now follow the same steps as in the Bloch reduction of Hamiltonian. The Hamiltonian eigenvectors are expressed as linear combinations
\begin{equation}\label{EGenSta}
|\psi_a(\Gamma)\rangle = \sum_{j}\sum_{A\in \mathcal{G}_{j}} c_{j}(\Gamma) \sqrt{s_j} |A\mathbf{0},A\mathbf{a}_j,a\rangle,
\end{equation}
where subset $\mathcal{G}_{\mathbf{0},j} \equiv \mathcal{G}_{j}\subseteq \mathcal{G}$ contains elements giving each atom $(A\mathbf{0},A\mathbf{a}_j)$ ones. Then, the Hamiltonian eigenvalue problem can be written as the set of homogenous equations in constants $c_{j}(\Gamma)$ with no summation over the repeated index $a$:
\begin{equation}\label{EHamRedHom}
\sum_{j'} \left[ \sum_{A\in \mathcal{G}_{j'},b} {D^{(\Gamma)}_{ab}}^*(A) H^{e,a;j}_{A,b;j'}(\Gamma) - E(\Gamma)\delta_{j,j'}\right] c_{j'}(\Gamma) = 0.
\end{equation}
In these equations $e$ is the unit group element and corresponds to $\mathbf{R}=\mathbf{0}$ in the Bloch theorem. This set of equations can be expressed as the eigenvalue problem
\begin{align}\label{EHamRed}
H(\Gamma) c(\Gamma) &= E(\Gamma) c(\Gamma)\\
H(\Gamma) &= \sum_{A\in \mathcal{G}_{j},b} {D^{(\Gamma)}_{ab}}^*(A) H^{e,a}_{A,b}(\Gamma),\nonumber
\end{align}
of the reduced Hamiltonian $H(\Gamma)$ and the matrix elements of $H^{e,a}_{A,b}(\Gamma)$ are
\begin{equation}\label{EHamRedMatEl}
H^{e,a;j}_{A,b;j'}(\Gamma) =
\sqrt{\frac{s_{j'}}{s_{j}}}\langle \mathbf{0},\mathbf{a}_{j},a | H \left( D_{j'}(A) | \mathbf{0},\mathbf{a}_{j'},b \rangle\right),
\end{equation}
where $D_{j'}(A) | \mathbf{0},\mathbf{a}_{j'},b \rangle$ is the orbital on the atom $(A\mathbf{0},A\mathbf{a}_{j'})$.
The index $a$ on the left-hand side in Eq.~(\ref{EHamRed}) is omitted as the reduced Hamiltonian does not depend on the choice for the nonzero vector $| \mathbf{0},\mathbf{a}_{j},a \rangle$.

In this section we represented the Hamiltonian in a symmetry adapted basis. Therefore the possible action of symmetries on the Hamiltonian can be made explicit and we can ask whether the Hamiltonian is invariant under an arbitrary action of the symmetries. To claim the invariance of the constructed Hamiltonian(s) we have to find and examine the structure of $|\mathbf{0}, \mathbf{a}_j \rangle\rangle$ vectors, defined by $P^{(\Gamma)}_{j}|\mathbf{0}, \mathbf{a}_j \rangle\rangle = |\mathbf{0}, \mathbf{a}_j \rangle\rangle$, and their transformation properties. We will do this in the following section.

\section{The symmetry constraint on gauge}
Let the symmetry group of a crystal have, beside translations ($\mathbf{R}$), one or more orthogonal transformations, $A$. An orthogonal transformation leaves the origin ($\mathbf{R}=\mathbf{0}$) intact. We will consider only the orthogonal transformations that do not commute with translations as only such transformations affect $\mathbf{k}$ and vectors $|\mathbf{0},\mathbf{a}_j\rangle\rangle$. Namely, it can happen that an orbit representative is not on the axis of rotation, and/or in the reflection plane, etc., and an additional translation is needed to bring the atom to its original position after some orthogonal transformation. In such situations, $\mathbf{k}$-dependant phases appear in vectors $|\mathbf{0},\mathbf{a}_j,a\rangle$. To extract this dependance, we need further analysis.

If the reduced projectors~(\ref{EgBtS}) are multidimensional ($\mathrm{Tr}P^{(\Gamma)}_j>1$), the results of the analysis coincide~\footnote{An example of this situation can be found in kagome lattice model in Sec. V.} with the case of equal numbers of orbit representatives and atoms in the unit cell, when the reduced projectors are one-dimensional. Therefore, the symmetry group will be reduced by the factor of maximal $\mathrm{Tr}P^{(\Gamma)}_j$, and all projectors will be one-dimensional. Notice that, in this case, the number of orthogonal transformations is equal to $d_{\Gamma} = s_j$ and all $s_j$ are same.

In the analysis of the $\mathbf{k}$-dependance of phases we will work only when $\Gamma$ is the general irreducible representation (label $\Gamma$ is redundant). Phases for the irreducible representations at special points and along special lines can be deduced as limits of the phases for a general IR. To find $\mathbf{k}$-dependance of phases we will explicitly construct IR using the method of induction. Induction of a (general) IR is performed, usually, from (Abelian) translational group with the one-dimensional IR $D^{(\mathbf{k})}(\mathbf{b}_i) = \exp( -\mathrm{i} \mathbf{k} \cdot \mathbf{b}_i )$. In this case, the IR of the orthogonal group transformations, $D(A)$, does not depend on the quasi-momentum and the matrix element (1,1) of the general $d_{\Gamma}$-dimensional IR for translation is $D_{11}(\mathbf{b}_i) = \exp( -\mathrm{i} \mathbf{k} \cdot \mathbf{b}_i ).$ The irreducible representation of translations is diagonal and other diagonal matrix elements satisfy
\begin{equation}\label{EdiagTIR}
\sum_{b=1}^{d_{\Gamma}}
D_{1b}(A) D_{bb}(\mathbf{b}_i) D_{b1}(A^{-1}) =
\mathrm{e}^{ -\mathrm{i} \mathbf{k} A \mathbf{b}_i }.
\end{equation}
To understand this requirement we notice that orthogonal transformations ($A$) when acting on vector $|\mathbf{0},\mathbf{a}_j\rangle\rangle$ rearrange its components i.e. partners of the multiplet. Given a vector $|\mathbf{0},\mathbf{a}_j\rangle\rangle$ we can track down which component ``$a$" of $|\mathbf{0},\mathbf{a}_j\rangle\rangle$ is transformed to the first component of $|A\mathbf{0},A\mathbf{a}_j\rangle\rangle$. In this way the particular $a$ and $A$ are connected in the way of a bijection which we denote by $A(a)$. Thus we have row: $D_{1b}(A)=\delta_{ab}$ and column $D_{b1}^{-1}(A)=\delta_{ab}$ in Eq.~(\ref{EdiagTIR}), i.e. the sum is redundant. At the end we see that the requirement in Eq.~(\ref{EdiagTIR}) means that the equivalent translation for $\mathbf{b}_i$ in the reference frame of the partner $a$ is given by $D_{aa}(\mathbf{b}_i) =\exp[-\mathrm{i} \mathbf{k} A(a) \mathbf{b}_i]$.

The physical content of orbitals is irrelevant in the analysis, and we will take representations $D_{j}$ to be one-dimensional identity operator. If two orbitals $(\mathbf{0},\mathbf{a}_j)$ and $(A\mathbf{0},A\mathbf{a}_j)$ are connected by the orthogonal group element $A$, then the vector $|A\mathbf{0},A\mathbf{a}_j\rangle\rangle$ is a linear combination of vector's $|\mathbf{0},\mathbf{a}_j\rangle\rangle$ components, see Eqs.~(\ref{EgBt}). On the other hand, there is a translation $\mathbf{R}$ that maps the initial atom $(\mathbf{0},\mathbf{a}_j)$ to $(A\mathbf{0},A\mathbf{a}_j)$ and each component $|\mathbf{0},\mathbf{a}_j,a\rangle$ gains the phase factor, see Eq.~(\ref{EdiagTIR}), $D_{aa}^*(\mathbf{R}) = \exp\{ \mathrm{i} \mathbf{k} A(a) \mathbf{R} \}$. The equivalence of these two symmetry operations implies a consistency condition that has to be imposed on the vectors $|\mathbf{0}, \mathbf{a}_j \rangle\rangle$. This (formally) coincides with the usual scalar condition in the quantum field theory: $\Psi'(\mathbf{k}) \equiv D(A) \Psi (\mathbf{k}) = \Psi (A^{-1}\mathbf{k})$, where $\Psi (\mathbf{k})$ fields in our case are vectors $|\mathbf{0}, \mathbf{a}_j \rangle\rangle$. The solution is given in the following general form:
\begin{equation}
|\mathbf{0}, \mathbf{a}_j, a \rangle(\mathbf{k}) = \exp\{ \mathrm{i} \mathbf{k} A(a)\mathbf{a}_j - \mathrm{i} \mathbf{k} \mathbf{a}_j^0 \} |\mathbf{0}, \mathbf{a}_j, a \rangle(\mathbf{k}=\mathbf{0}).
\label{solution}
\end{equation}
The phase part affected by orthogonal transformations is fixed by the consistency requirement. If an orthogonal transformation ($A$) transfers vector $\mathbf{a}_j$ into vector $A\mathbf{a}_j$ that is equal to $\mathbf{a}_j+\mathbf{R}$, where $\mathbf{R}$ is the corresponding translation, i.e. $A\mathbf{a}_j=\mathbf{a}_j+\mathbf{R}$, and the diagonal matrix of the translation is $D_{aa}^*(\mathbf{R}) = \exp\{ \mathrm{i} \mathbf{k} A(a) \mathbf{R} \}$, (see Eq.~(\ref{EdiagTIR}) and below) the phase part must be $\exp\{ \mathrm{i} \mathbf{k} A(a)\mathbf{a}_j\}$. There is also an overall phase freedom given by the phase, $\exp\{ - \mathrm{i} \mathbf{k} \cdot \mathbf{a}_j^0 \}$, where $\mathbf{a}_j^0$ is invariant under orthogonal transformations. This invariance of $\mathbf{a}_j^0$ under orthogonal transformations ensures consistent definition (the same gauge) for the orbitals connected by the group action.

Let as first discuss the case of isotropic interactions. Isotropic interactions are defined by the condition that the strength between a fixed orbital $j$ and all $j'$ at the same distance from $j$ is the same. Then the only $\mathbf{a}_j^0$ invariant under all orthogonal transformations is the zero vector, i.e. $\mathbf{a}_j^0 \equiv \mathbf{0}$. Any orthogonal transformation will act only as an orthogonal matrix. Under an arbitrary orthogonal transformation $B$ on the $|\mathbf{0}, \mathbf{a}_j \rangle\rangle$ vectors in the generalized Bloch Hamiltonian, Eq.~(\ref{EHamRedMatEl}), the summation over $A$ becomes the one over $B^{-1} A B = A'$, which covers again ${\cal G}_j$, and makes the Hamiltonian invariant.

The unique Hamiltonian defined by $\mathbf{a}_j^0 \equiv \mathbf{0} $ in this generalization of the Bloch theorem coincides with the Hamiltonian in the periodic gauge. The periodic gauge is defined as the gauge in which the Bloch vector, $|\psi(\mathbf{k})\rangle$, is invariant under the translation by $ \mathbf{G} = \sum_{i = 1}^{d} m_i \mathbf{g}_i$ in the inverse space. Then the coefficients in Eq.~(\ref{EGenSta}) $c_j(\mathbf{k})$ transform as:
\begin{equation}
c_j(\mathbf{k}+ \mathbf{G}) = c_j(\mathbf{k}) \exp\{- \mathrm{i} \mathbf{G} \mathbf{a}_j\}.
\label{trans}
\end{equation}
In the case when $\mathbf{a}_j^0 \equiv \mathbf{0} $ for each $j$ the Hamiltonian matrix elements transform as
\begin{equation}
H^{a; j}_{b; j'} (\mathbf{k}+ \mathbf{G}) = H^{a; j}_{b; j'} (\mathbf{k}) \exp\{ \mathrm{i} \mathbf{G} (\mathbf{a}_j - \mathbf{a}_{j'} )\}.
\end{equation}
This follows from the solution (\ref{solution}) for the vectors that define the generalized Bloch Hamiltonian and the fact that, in each component of the vector, $A(a) \mathbf{a}_j$ in the exponent can be rewritten as $ A(a) \mathbf{a}_j = \mathbf{R} + \mathbf{a}_j$, i.e. as a translation by $ \mathbf{R}$, a Bravais lattice vector. The invariance of the eigenvalue problem under the translation by $ \mathbf{G}$ implies (\ref{trans}).
Thus, in the case of isotropic interactions, the generalized Bloch Hamiltonian is the unique Hamiltonian that is invariant under orthogonal transformations and coincides with the Hamiltonian in the periodic gauge.

We showed that the requirement that the Bloch eigenvector is invariant under the translation by any $ \mathbf{G}$ in the inverse space, $|\psi( \mathbf{k} + \mathbf{G}) \rangle = |\psi( \mathbf{k}) \rangle$, i.e. the requirement on its coefficients in (\ref{trans}), or the requirement that the Hamiltonian is invariant under orthogonal transformations lead to the same Hamiltonian.

Let us now consider anisotropic interactions. In this case the invariance under orthogonal transformations does not constrain all $\mathbf{a}_j^0$ to zero, i.e. there will be non-zero $\mathbf{a}_j^0$ which are invariant under orthogonal transformations. Thus instead of a unique Hamiltonian we can have a family of Hamiltonians invariant under orthogonal transformations. The Hamiltonian that represents the periodic gauge choice, i.e.  $\mathbf{a}_j^0 \equiv \mathbf{0} $ for each $j$, belongs to this family. The anisotropic case may be considered as an extension of the isotropic case when we start to continuously change the interaction strength along some directions. If we demand a continuous change of Berry curvature, or any physical quantity in the inverse space, along this process, then the periodic gauge choice is a natural choice even in the case of anisotropic interactions.

We see that in the case of the anisotropic interactions the generalized Bloch theorem, which ensures the invariance under orthogonal transformations does allow various gauges with some of $\mathbf{a}_j^0$'s being non-zero. If furthermore the choice of $\mathbf{a}_j^0$'s leads to periodicity of the coefficients, $c_j(\mathbf{k})$, $c_j(\mathbf{k}+ \mathbf{G}) = c_j(\mathbf{k})$, and of the Hamiltonian, we deal with the Bloch Hamiltonians. The periodicity in the inverse space is the usual, defining requirement for the Bloch Hamiltonians. This choice for the Hamiltonian, so-called Bloch gauge choice is often used in the literature on the fractional Chern insulators, like in Refs.~\cite{PhysRevLett.106.236804,PhysRevB.85.075116,PhysRevB.85.075128}, irrespective whether the Bloch Hamiltonian is invariant under orthogonal transformations. As long as we ask for physical (``gauge invariant") quantities, like energy spectrum, degeneracy of the ground state on the torus, Chern number, this is justified. But if we seek description in the inverse space and look for quantities that may be considered as observables, like density in the inverse space, internal field, i.e. Berry curvature or the quantities that depend on local values of Berry curvature like anomalous Hall conductance we have to be careful. We can find the quantities (easily) in the periodic gauge (the natural gauge that respects symmetries) or if we use Bloch gauge we have to keep track and include real space embeddings of orbitals~\cite{PhysRevB.86.085129}.

As we already emphasized there is the need for the periodic gauge even in the case of anisotropic interactions where the Bloch Hamiltonian may be invariant under orthogonal transformations. In general the off-diagonal elements of the Bloch Hamiltonians under orthogonal transformations acquire phase factors that depend on the quasi-momentum $\mathbf{k}$ - see Table~\ref{Tbloper}. But though the periodic gauge is unique and physical, as we demonstrated by developing the generalized Bloch theorem, there is an interesting connection between the Berry curvatures of all possible Bloch Hamiltonians and the Berry curvature in the periodic gauge. Namely, as we will demonstrate in examples, the average of the Berry curvatures of Bloch Hamiltonians, defined by a choice of the unit cell and its partner cells connected by orthogonal transformations, is equal to the Berry curvature in the periodic gauge.
\begin{table}[htbp]
  \centering
  \caption{Phases $\varphi(\mathbf{k})$ of the phase factor $\exp[\mathrm{i}\varphi(\mathbf{k})]$ that could be acquired by the certain transformations of quasi-momentum $\mathbf{k}$ in the off-diagonal Hamiltonian matrix elements in two gauges. Diagonal elements do not change.}\label{Tbloper}
  \begin{tabular}{c|c|c|c|}
    \multicolumn{2}{c|}{} & \multicolumn{2}{c|}{Transformation} \\\cline{3-4}
    \multicolumn{2}{c|}{} & Orthogonal & Adding $\mathbf{G}$ \\\hline
    \multirow{2}{*}{Gauge} & Bloch & $\varphi(\mathbf{k})$ & 0 \\\cline{2-4}
     & Periodic & 0 & const. \\\hline
  \end{tabular}
\end{table}

\section{Berry curvature characterization}
In the following two-dimensional examples, quasi-momentum is parameterized as $\mathbf{k} = (k_1,k_2) = ( \mathbf{k}\cdot\mathbf{b}_1, \mathbf{k}\cdot\mathbf{b}_2 )$. As we are interested in the Berry curvature, IRs are general and $D_{j}=1$. Berry curvature is defined as
\begin{equation}\label{EBerrycurv}
B=\mathrm{i}( \langle\partial_{k_x}u_n(\mathbf{k})|\partial_{k_y}u_n(\mathbf{k})\rangle - \langle\partial_{k_y}u_n(\mathbf{k})|\partial_{k_x}u_n(\mathbf{k})\rangle ),
\end{equation}
where $|u_n(\mathbf{k})\rangle$ is the Hamiltonian eigenvector, $n$ is the band index, and derivatives are along $k_x$ and $k_y$ direction in reciprocal space. Besides $U(1)$ symmetry that leaves Berry curvature unchanged, permutation of vector's $|u_n(\mathbf{k})\rangle$ components also leaves invariant Berry curvature. These permutations are, in fact, freedom in the choice of the order of the basis vectors $|\mathbf{0},\mathbf{a}_j\rangle.$

All three models, analyzed here, are relevant in the description of (fractional) quantum anomalous Hall effect, i.e. (fractional) Chern insulators. The Berry curvature, usually viewed as an intrinsic magnetic field in reciprocal space, is responsible for various effects. The dispersion of its local values gives a criterion for the stability of fractional Chern insulator states~\cite{PhysRevB.85.241308,PhysRevB.85.075116,PhysRevB.85.115117,arXiv.1408.0843}. Naturally, its symmetry has to be the same as the symmetry of a crystal. In the examples in this section it is shown that Hamiltonian constructed by the generalized Bloch theorem, i.e. by using full symmetry of a crystal, have the most symmetric Berry curvature. In addition, this curvature is compared with the curvatures of the Bloch Hamiltonians.

We will give detailed explanations in the first example - the Haldane model, making its exposition self-contained and also enabling a reader to follow easily the rest of the examples in this section.

\begin{figure}[hbt]\centering
\includegraphics[width=0.5\linewidth]{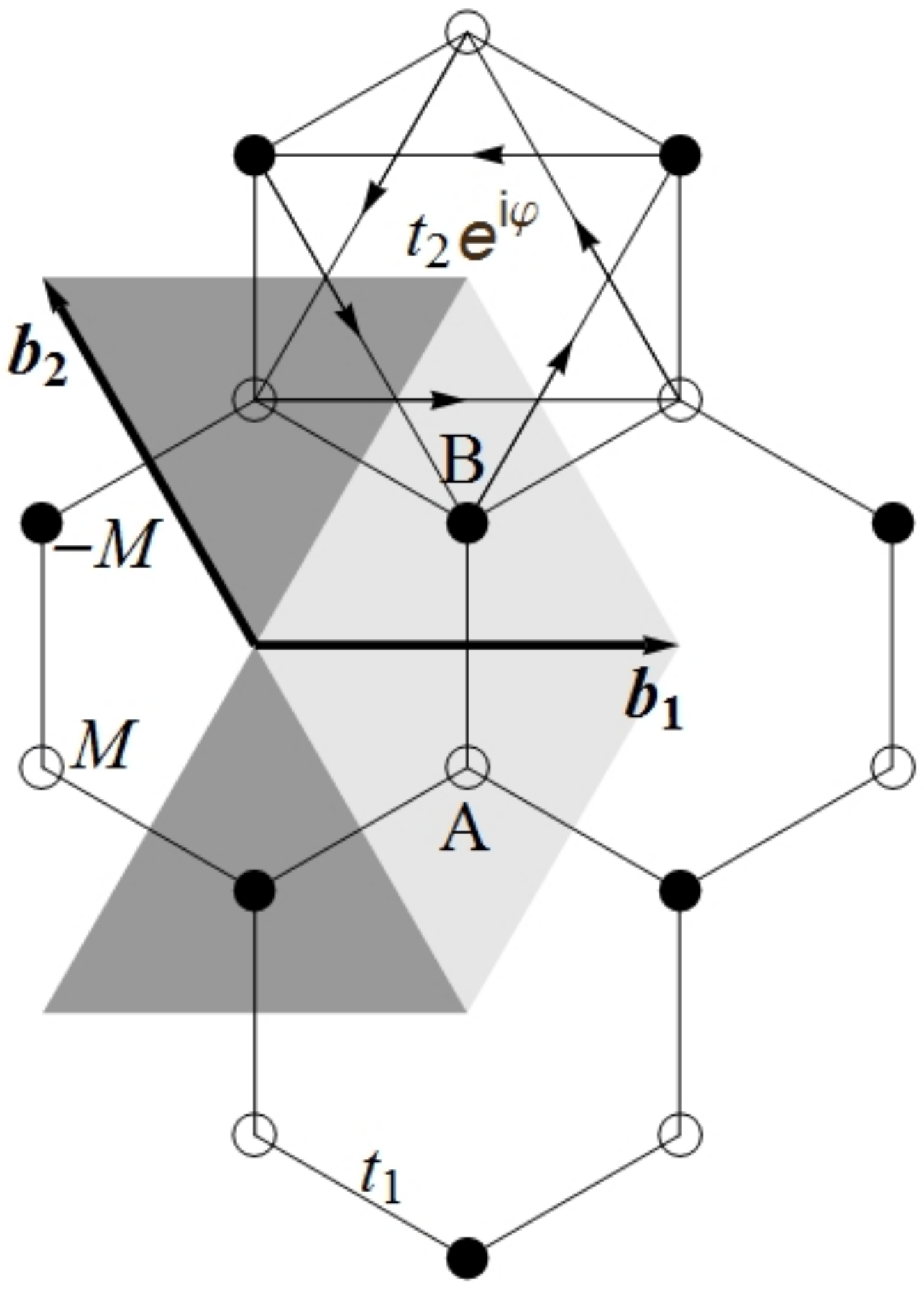}
\caption{\label{FHaldane} Honeycomb lattice with two sublattices $A$ (circles) and $B$ (disks), with on-site energies $M$ and $-M$, respectively. Nearest neighbors hoping amplitude is $t_1$, while next nearest neighbor complex hoping amplitude is $t_2\mathrm{e}^{\mathrm{i}\varphi}$ in the arrow direction. Vectors $\mathbf{b}_1$ and $\mathbf{b}_2$ are direct lattice vectors. Light and dark grey areas are two differently shaped unit cells.}
\end{figure}

The first and most relevant example is Haldane model~\cite{Haldane-PhysRevLett.61.2015} for honeycomb lattice (see Fig.~\ref{FHaldane}).
The symmetry group of this model is $\mathcal{C}_{3h}\mathcal{T}_2$, where $\mathcal{C}_{nh}$ (in Schoenflies notation) is the group of rotations for $2\pi/n$ about $z$-axis and reflection in horizontal plane, $\mathcal{T}_2$ is the group of two-dimensional translations along vectors $\mathbf{b}_1=a\sqrt{3}(1,0)$ and $\mathbf{b}_2=a\sqrt{3}(-1/2,\sqrt{3}/2)$, and $a$ is the nearest neighbor (NN) distance. Here, for simplicity, the symmetry group is taken to be $\mathcal{C}_{3}\mathcal{T}_2$.

In the following, we will in simple terms, illustrating the general approach described in Sections III and IV, explain the steps that lead to a unique (generalized Bloch) Hamiltonian that respects all symmetries. The first step is to find vectors on which the symmetry is realized, i.e. on which the irreducible representation acts on and realizes the symmetry operations. The second step is to use these vectors to constrain the space of the eigenproblem - these vectors will make a (symmetry constrained) basis for the eigenproblem, and express the Hamiltonian in this basis.

To start with, let us consider all symmetry operations and objects (general vectors) that they act on in the case of the Haldane model. In Fig.~\ref{FHaldane} we choose light gray area as a unit cell, and the origin is fixed at the beginning of vectors $\mathbf{b}_1$ and $\mathbf{b}_2$. The symmetry operation, $C_3$, counterclockwise rotation for $2\pi/3$ around the origin, acts as a permutation, $1\rightarrow2\rightarrow3\rightarrow1$, for three $A$ or $B$ atoms. So, besides considering spacial images like $C_3A$, we may associate 3-tuples to each atom and consider that 3-tuple at $A$ is transformed into 3-tuple at $C_3A$ by the following matrix that represents permutation:
\begin{align}
D(C_3) &= \begin{pmatrix} 0&0&1\\1&0&0\\0&1&0\end{pmatrix}.\nonumber
\end{align}
This represents the part, i.e. the generator in the IR corresponding to rotations in the group $\mathcal{C}_3\mathcal{T}_2$. To construct the generators $D(\mathbf{b}_1)$ and $D(\mathbf{b}_2)$, for the translations $\mathcal{T}_2$ we first notice that we may associate the first component of 3-tuple at atom $A$ in the unit cell to the description of that atom $A$ in that cell because of the permutational link among 3-tuples of three $A$ atoms connected by rotation. Thus, the first component of 3-tuple at $A$ (that belongs to the unit cell) should transform by an ordinary translation, i.e. should be multiplied by $\exp(-\mathrm{i}k_1)$ and $\exp(-\mathrm{i}k_2)$ for the translations by $\mathbf{b}_1$ and $\mathbf{b}_2$ respectively. We reach the rest of the atoms by rotations. It is natural to represent equivalent translations in their reference frames as $\exp(-\mathrm{i}\mathbf{k}R\mathbf{b}_1)$ and $\exp(-\mathrm{i}\mathbf{k}R\mathbf{b}_2)$, where $R$ is the orthogonal transformation, i.e. $R=C_3^{-1}, C_3^{-2}$ rotation that is necessary to reach the reference frames of the other two atoms. In this way we avoid singling out atom $A$ in the unit cell and consider all three atoms at equal footing, which is a symmetry requirement. Thus the generators for the translations are:
\begin{align}
D(\mathbf{b}_1) &= \mathrm{diag}(\mathrm{e}^{-\mathrm{i}k_1} , \mathrm{e}^{\mathrm{i}(k_1+k_2)} , \mathrm{e}^{-\mathrm{i}k_2}),\nonumber\cr
D(\mathbf{b}_2) &= \mathrm{diag}(\mathrm{e}^{-\mathrm{i}k_2} , \mathrm{e}^{-\mathrm{i}k_1} , \mathrm{e}^{\mathrm{i}(k_1+k_2)} ).\nonumber
\end{align}
This concludes the illustration in the Haldane case of the general method of the induction of IR generators given by Eq.~(\ref{EdiagTIR}) in Sec.~IV.

Then we consider atoms $A(a\sqrt{3}/2,-a/2)$ and $B(a\sqrt{3}/2,a/2)$ (see Fig.~\ref{FHaldane}) in the unit cell as two orbit representatives, i.e. representatives of two groups of atoms that cannot be connected by symmetry operations. We call the stabilizers of orbit representatives, $A$ and $B$, two subgroups of the symmetry group, $S_A$ and $S_B$, that leave the positions of atom $A$ and atom $B$ unchanged, respectively. They are
\begin{align}
S_A &= \{ e , ( C_3 | -\mathbf{b}_2 ) , ( C_3^2 | \mathbf{b}_1 ) \},\nonumber\cr
S_B &= \{ e , ( C_3 | \mathbf{b}_1 ) , ( C_3^2 | \mathbf{b}_1 + \mathbf{b}_2 ) \},\nonumber
\end{align}
where $e\equiv (e | \mathbf{0})$ is the identity group element and $(R|\mathbf{R})$ is the group element in Koster-Seitz notation (orthogonal transformation $R$ followed by translation for vector $\mathbf{R}$). The 3-tuples at atoms $A$ and $B$ we denote by $|A\rangle\rangle$ and $|B\rangle\rangle$ respectively. To fulfill the requirement that the symmetry is realized on these 3-tuples, we demand that all members of $S_A$, which we denote by $\bar{A}$ leave $|A\rangle\rangle$ unchanged, i.e.
\begin{equation}\label{EHalProj}
|A\rangle\rangle = \frac13\sum_{\bar{A}\in S_{A}}D^*(\bar{A}) |A\rangle\rangle = P_A |A\rangle\rangle
\end{equation}
We applied Eq.~(\ref{EgBtT}) for each symmetry operation and summed.
In Eq.~(\ref{EHalProj}) $P_A$ denotes a projector ($P_A^2=P_A$ and $P_A^{\dagger} = P_A$) that constrains the description of $|A\rangle\rangle$, i.e. singles out a certain combination of orbitals. Analogously we place a constraint on vector $|B\rangle\rangle$, i.e. $|B\rangle\rangle = P_B |B\rangle\rangle.$ Projectors $P_A$ and $P_B$ correspond to the ones that enter Eq.~(\ref{EgBtS}) in Sec.~IV of the general construction. These projectors are one-dimensional, i.e. $\mathrm{Tr}P_A =\mathrm{Tr}P_B =1$, and determine $|A\rangle\rangle$ and $|B\rangle\rangle$ up to a phase. Nevertheless, we can infer what that phase should be. Namely, as can be found from projectors $P_A$ and $P_B$ or from a general consistency argument between translations and rotations in Sec.~IV, see Eq.~(\ref{solution}), the solutions can be expressed in the following form:
$$
|X, a \rangle(\mathbf{k}) = \exp\{ \mathrm{i} \mathbf{k} R_X(a)\mathbf{a}_X - \mathrm{i} \mathbf{k} \mathbf{a}_X^0 \} |X, a \rangle(\mathbf{k}=\mathbf{0}),
$$
where $X=A,B$, and $a=1,2,3$ for the three components of the IR. Also $R_X(a)$ can be understood as the symmetry operation to transfer atom $X$ from the unit cell to the positional partner $a$ where we have $R_X(1)=I.$ Vectors $\mathbf{a}_{A/B}$ in the phases of both $|A,1\rangle$ and $|B,1\rangle$ are positions of the orbit representatives, while the vectors in the second and third components are position of atoms $C_3^{-1} \mathbf{a}_{A/B}$ and $C_3^{-2} \mathbf{a}_{A/B}$, respectively. The overall phase with vector $\mathbf{a}_X^0$ is constrained by the requirement that $\mathbf{a}_X^0$ is invariant under all orthogonal transformations. This ensures the same gauge i.e. a consistent definition - fixing of 3-tuples in whole lattice space, i.e. atoms that can be reached by orthogonal transformations $R: R\mathbf{a}_X^0=\mathbf{a}_X^0.$ In this way only the part with $\mathbf{a}_X$ is affected, i.e. $\mathbf{a}_X\rightarrow R\mathbf{a}_X.$ In this example the introduced vector $\mathbf{a}_X^0$ is zero vector, as the only vector in the $xy$-plane invariant under the rotations. In three dimensions this vector can be along $z$-axis, but by adding horizontal plane reflection it becomes zero. Thus
\begin{align}
|A\rangle\rangle &=( \mathrm{e}^{\mathrm{i}\frac{k_1-k_2}{3}} , \mathrm{e}^{-\mathrm{i}\frac{2k_1+k_2}{3}} , \mathrm{e}^{\mathrm{i}\frac{k_1+2k_2}{3}} )^{\mathrm{T}},\nonumber\cr
|B\rangle\rangle &=( \mathrm{e}^{\mathrm{i}\frac{2k_1+k_2}{3}} , \mathrm{e}^{-\mathrm{i}\frac{k_1+2k_2}{3}} , \mathrm{e}^{-\mathrm{i}\frac{k_1-k_2}{3}} )^{\mathrm{T}},\nonumber
\end{align}
where $\mathrm{T}$ stands for transposition.

Therefore we determined the two vectors of the orbit representatives, $|A\rangle\rangle \equiv |\mathbf{0},\mathbf{a}_A\rangle\rangle$ and $|B\rangle\rangle \equiv |\mathbf{0},\mathbf{a}_B\rangle\rangle$ and by that we determined - fixed all 3-tuples on all atoms of the lattice that can be reached by symmetry operations, $R$. These 3-tuples we denote by $|R\mathbf{0},R\mathbf{a}_X\rangle\rangle,$ $X=A,B.$

The next step, after the fixing the 3-tuples, i.e. the vectors on which the symmetry of the lattice is realized, is to look for the form of the Hamiltonian in the basis that these vectors make. An arbitrary vector from the constrained space or the one on which the eigenproblem is now defined, can be expressed as
$$
|\psi_a\rangle = \sum_{X,R\in \mathcal{G}_X} f_X |R\mathbf{0},R\mathbf{a}_X,a\rangle = \sum_X f_X \widehat{|X,a\rangle},
$$
where $f_X$ are the coefficients of the expansion.
In the previous equation $\mathcal{G}_X$ denotes set of group elements that connect atoms in orbit $X$. The label $a$ can denote any index of the IR - due to their equal roles i.e. symmetry, the choice of the index $a$ is arbitrary i.e. $a$ can be 1, 2, or 3 in our case. The final form of the Hamiltonian matrix does not depend on this choice. To find the form of the Hamiltonian in this space we have to evaluate the matrix elements $H^X_Y$, where $X$ and $Y$ can be either $A$ or $B.$ The matrix element $H^X_Y$ may be viewed as a sandwich between $\widehat{\langle X,a|}$ and $\widehat{|Y,a\rangle}$. But, because of the symmetry instead of $\widehat{\langle X,a|}$ we may take just $\langle X,a| \equiv \langle \mathbf{0},\mathbf{a}_X,a|$. In other words it is irrelevant which atom in orbit $X$ we choose as a reference point from which we will measure relevant interaction parameters for the matrix element $H^X_Y$. Thus, also following the general exposition in Eqs.~(\ref{EHamRed}) and~(\ref{EHamRedMatEl}), in this case in which $s_A=s_B=3,$ we have
$$
H^X_Y = \langle \mathbf{0},\mathbf{a}_X,a| H \widehat{|Y,a\rangle} = \sum_{R\in\mathcal{G}_Y} \langle \mathbf{0},\mathbf{a}_X,a| H |R\mathbf{0},R\mathbf{a}_Y,a\rangle.
$$
After an application of Eq.~(\ref{EgBtT}) (as a basic representation of group action) we have
\begin{equation}\label{EHaldOff}
H^X_Y = \sum_{b,R\in\mathcal{G}_Y}D^{*}_{ab}(R) \langle \mathbf{0},\mathbf{a}_X,a| H |\mathbf{0},\mathbf{a}_Y,b\rangle.
\end{equation}
This equation corresponds to Eqs.~(\ref{EHamRed}) and~(\ref{EHamRedMatEl}) in the general construction. For example, if $X=A$ and $Y=B$ we have to take into account the near-neighbor coupling $t_1$, between atom $A$ and atoms from the orbit with the representative $B$. Thus besides the sum over $b$, in order to evaluate $H^A_B$, we also have to sum over surrounding atoms that we enumerate by group elements $R$ necessary to reach them from the representative $B$, $\{e, C_3^2, -\mathbf{b}_2\}$.

As we pointed out in the previous paragraph, off-diagonal element $H^A_B$ describes the interaction of the orbit representative $A$ with its three nearest neighboring $B$ atoms. These three atoms are images of the group elements' $\{e, C_3^2, -\mathbf{b}_2\}$ action on the orbit representative $B$. Since the first component $|A,1\rangle = \exp[\mathrm{i}(k_1-k_2)/3]$ is non-vanishing, we can take $a=1$ in Eq.~(\ref{EHaldOff}) in this section, or Eqs.~(\ref{EHamRed}) and~(\ref{EHamRedMatEl}) in the general construction. Now, we need matrix elements of IR $D_{1b}(R)$, conjugate them, multiply with corresponding $|B,b\rangle$, and sum over $b$. Only non-zero matrix elements of IR for previous group elements are $D_{11}(e)=1$, $D_{12}(C_3^2)=1$, and $D_{11}(-\mathbf{b}_2)=\exp(\mathrm{i}k_2)$, and the Hamiltonian (NN interactions is $t_1$) off-diagonal matrix element is
\begin{align}
&\langle A,1| t_1\left[ \left(D^*_{11}(e) + D^*_{11}(-\mathbf{b}_2)\right) |B,1\rangle + D^*_{12}(C_3^2) |B,2\rangle \right] =\nonumber\\
&H^A_B = t_1 \left( \mathrm{e}^{\mathrm{i}\frac{k_1+2k_2}{3}} + \mathrm{e}^{\mathrm{i}\frac{k_1-k_2}{3}} + \mathrm{e}^{-\mathrm{i}\frac{2k_1+k_2}{3}} \right).\nonumber
\end{align}
Element $H^B_A$ is equal to ${H^A_B}^*$. To find diagonal matrix elements we have to sum over six interactions. Note that orbit representatives hoping amplitude with three atoms is $t_2\exp(\mathrm{i}\varphi)$ and with the other three $t_2\exp(-\mathrm{i}\varphi)$. In this way we obtain
$$
H_{A/B}^{A/B} = \pm M + $$$$2t_2[ \cos(k_1 \pm \varphi) + \cos(k_2 \pm \varphi) + \cos(k_1 + k_2 \mp \varphi) ],
$$
where upper signs are for $H_A^A$ and the lower ones for $H_B^B$.
Although stabilizers are different for atoms in the dark grey area (see Fig.~\ref{FHaldane}), we find that vectors $|A/B\rangle\rangle$ are just permutations of the previous vectors, and Hamiltonian matrix elements are the same. The resulting Hamiltonian stays the same for any choice of orbit representatives, not necessarily in a unit cell. Adding general reciprocal lattice vector $\mathbf{G} = m_1\mathbf{g}_1 + m_2\mathbf{g}_2$ to $\mathbf{k}$, the diagonal Hamiltonian elements stay as they are, while off-diagonal element $H^A_B$ gains constant (as $m_i$ are integers) phase $2\pi(2m_1+m_2)/3$. This does not affect the Berry curvature. The rotation of $\mathbf{k}$ vector for integer multiple of $2\pi/3$ does not change the Hamiltonian at all.

\begin{figure}[hbt]\centering
\includegraphics[width=0.7\linewidth]{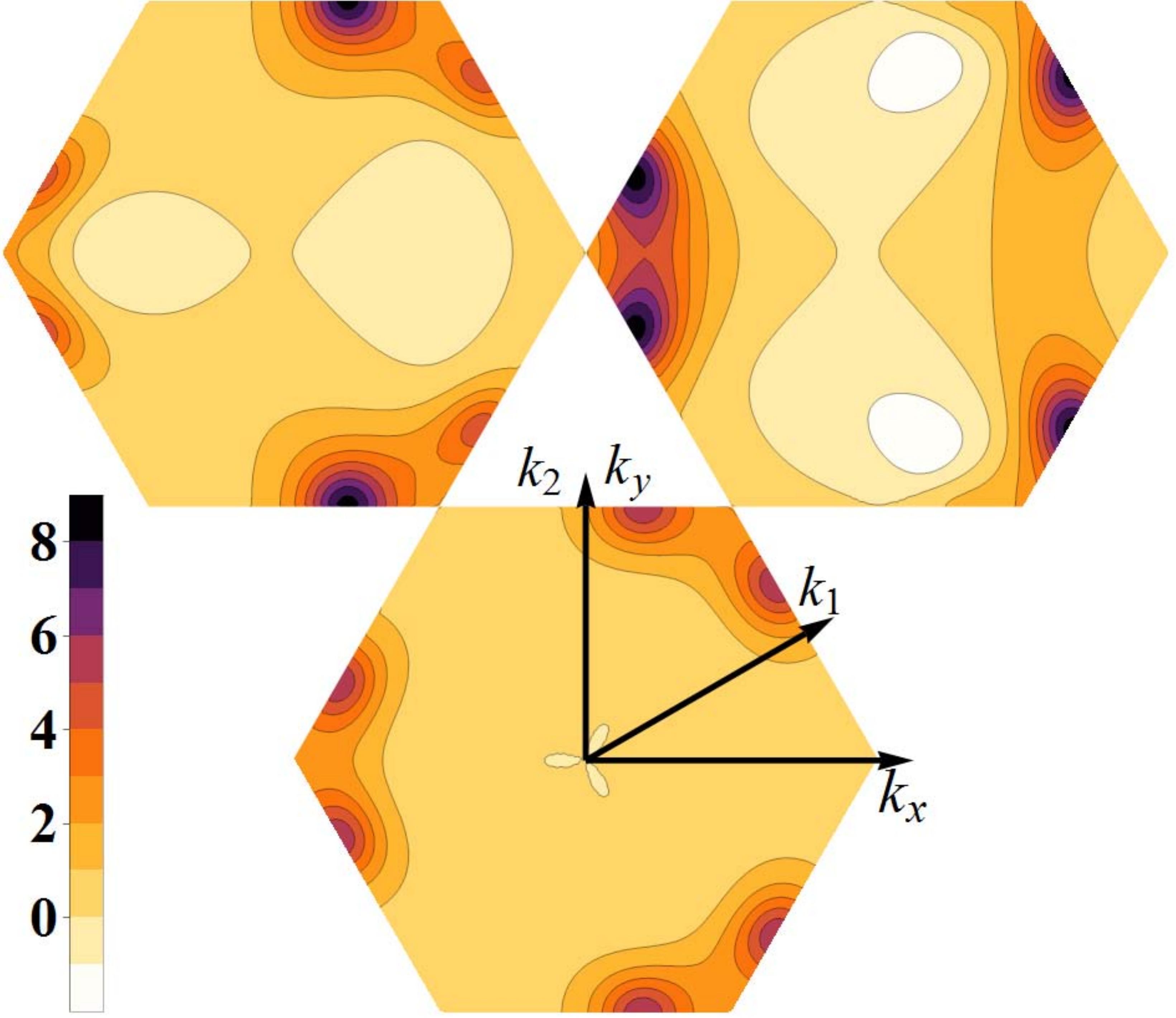}
\caption{\label{FHBerry} (Color online) Berry curvatures, in the Brillouin zone, of the lowest energy band for the Haldane model, with parameters $M=1$, $t_1=t_2=1$, and $\varphi=0.125\pi$. In the lower part is the curvature for the generalized Bloch Hamiltonian constructed here (Hamiltonian in the periodic gauge). Up-left is for the Bloch Hamiltonian for the light-grey (see Fig.~\ref{FHaldane}) and up-right for the dark-grey unit cell.}
\end{figure}

The Berry curvature of the lowest energy band for the Haldane model is given in Fig.~\ref{FHBerry}, parameters are $M=1$, $t_1=t_2=1$, and $\varphi=0.125\pi$.
It can be seen that the generalized Bloch Hamiltonian has the most symmetric Berry curvature. The curvatures of the Bloch Hamiltonians for the light and dark grey unit cells (see Fig.~\ref{FHaldane}) have only translational symmetry, in the reciprocal space. By rotating light and dark grey unit cells for $2\pi/3$ and $4\pi/3$ and constructing the Bloch Hamiltonians for such unit cells, Berry curvature is also rotated for the same angle. Average of all three Berry curvatures, separately for light and dark grey unit cells, is equal to the Berry curvature of the generalized Bloch Hamiltonian. Equivalent conclusion will hold in other examples. Notice that rotation of the light-grey unit cell is equivalent to fixing one atom and making the Bloch Hamiltonians for that atom and its nearest neighbors.

\begin{figure}[hbt]\centering
\includegraphics[width=0.7\linewidth]{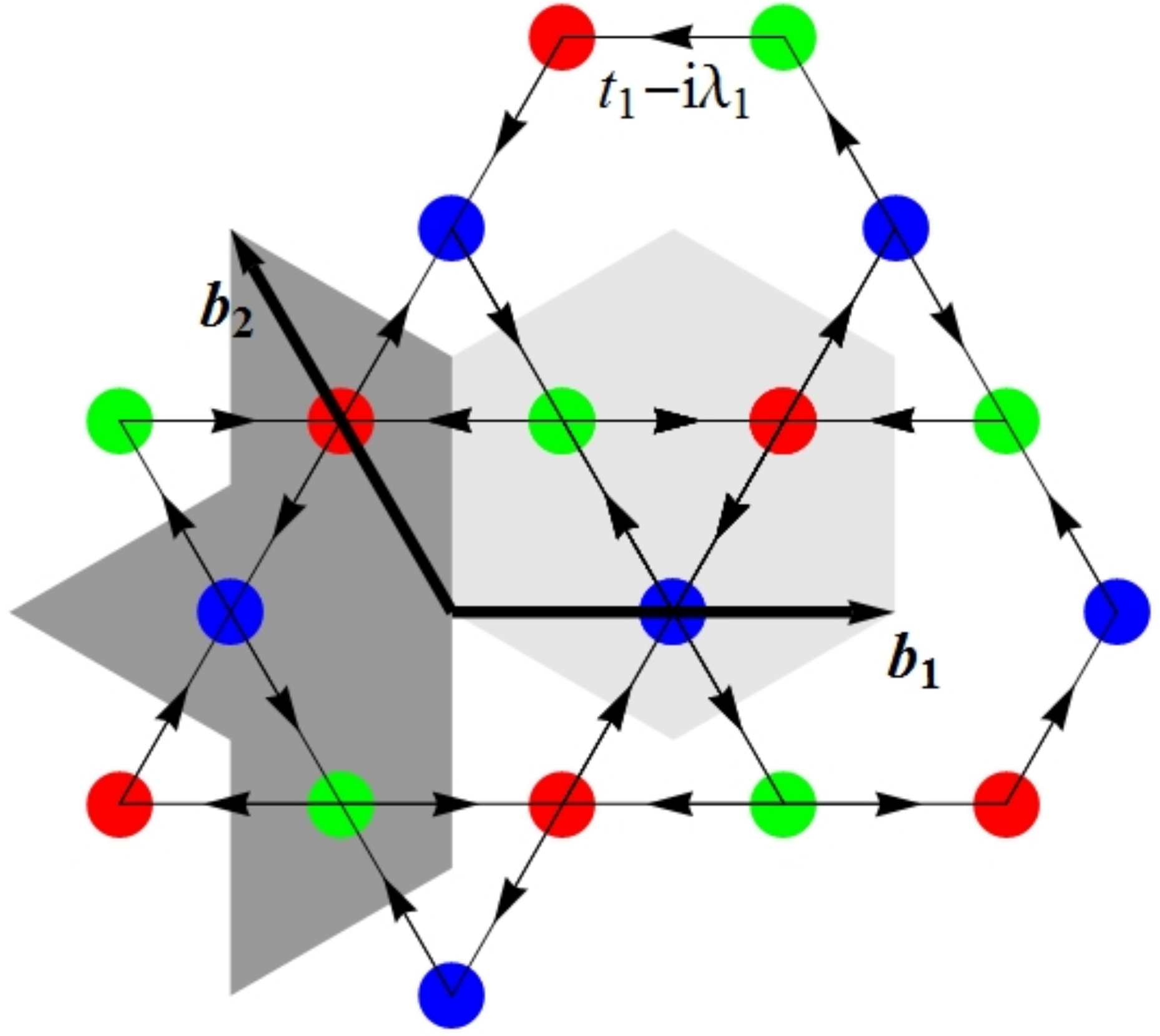}
\caption{\label{Fkagome} (Color online) Kagome lattice model with three sublattices $R$ [grey (red) disks], $G$ [light-gray (green) disks], and $B$ [dark-grey (blue) disks]. Complex nearest neighbors hoping amplitude is $t_1-\mathrm{i}\lambda_1 = t_1 \mathrm{e}^{-\mathrm{i}\phi}$ in the arrow direction. Vectors $\mathbf{b}_1$ and $\mathbf{b}_2$ are direct lattice vectors. Light and dark grey areas are two differently shaped unit cells.}
\end{figure}

The second example is kagome~\cite{PhysRevLett.106.236802,Guocai-PhysRevB.79.035323} lattice model (see Fig.~\ref{Fkagome}).
The atoms are indistinguishable, and the symmetry of the crystal is a product of axial group $\mathcal{C}_6$ and 2D translations along $\mathbf{b}_1=2a(1,0)$ and $\mathbf{b}_2=2a(-1/2,\sqrt{3}/2)$, where $a$ is the nearest neighbor distance. The whole crystal can be obtained by this group action on a single atom. Nevertheless, the resulting Hamiltonian is equal to the Hamiltonian constructed with symmetry group $\mathcal{C}_2\mathcal{T}_2$ having three orbit representatives. For the sake of simplicity and to have the representation by vectors with lower dimension (2 instead of 6), we will take as symmetry group $\mathcal{C}_2\mathcal{T}_2$ with three orbit representatives. The general irreducible representation of the generators is:
\begin{align}
D(C_2) &= \begin{pmatrix} 0&1\\1&0\end{pmatrix},\nonumber\cr
D(\mathbf{b}_1) &= \mathrm{diag}(\mathrm{e}^{-\mathrm{i}k_1}, \mathrm{e}^{\mathrm{i}k_1}),\nonumber\cr
D(\mathbf{b}_2) &= \mathrm{diag}(\mathrm{e}^{-\mathrm{i}k_2}, \mathrm{e}^{\mathrm{i}k_2}).\nonumber
\end{align}
As the resulting Hamiltonian does not depend on the choice of orbit representatives, we will take, for example, atoms from the dark-grey unit cell $R(-a/2,a\sqrt{3}/2)$, $G(-a/2,-a\sqrt{3}/2)$, and $B(-a,0)$ (see Fig.~\ref{Fkagome}). Stabilizers for these atoms are $S_R = \{ e, (C_2|\mathbf{b}_2) \}$, $S_G = \{ e, (C_2|-\mathbf{b}_1-\mathbf{b}_2) \}$, and $S_B = \{ e, (C_2|-\mathbf{b}_1) \}$. Again, the projectors for these stabilizers are one-dimensional with vectors
$$
|R\rangle\rangle =\begin{pmatrix} \mathrm{e}^{\mathrm{i}\frac{k_2}{2}} \\ \mathrm{e}^{-\mathrm{i}\frac{k_2}{2}} \end{pmatrix},
|G\rangle\rangle =\begin{pmatrix} \mathrm{e}^{-\mathrm{i}\frac{k_1+k_2}{2}} \\ \mathrm{e}^{\mathrm{i}\frac{k_1+k_2}{2}} \end{pmatrix},
|B\rangle\rangle =\begin{pmatrix} \mathrm{e}^{-\mathrm{i}\frac{k_1}{2}} \\ \mathrm{e}^{\mathrm{i}\frac{k_1}{2}} \end{pmatrix}
$$
in their ranges and in the necessary gauge. Here, as in the previous example, vectors $\mathbf{a}_{j}^0$ are zero. Also, the first components have $\mathbf{a}_{j}$ vectors equal to the positions of orbit representatives, while, in the second components, vectors are the positions of atoms rotated for $\pi$.

Now, as in the Haldane example, with the help of Eqs.~(\ref{EHamRed}) and~(\ref{EHamRedMatEl}), the generalized Bloch Hamiltonian is
$$
H = 2t_1
\begin{pmatrix}
0 & \mathrm{e}^{-\mathrm{i}\phi}\cos\frac{k_1}{2} & \mathrm{e}^{\mathrm{i}\phi}\cos\frac{k_1+k_2}{2} \\
\mathrm{e}^{\mathrm{i}\phi}\cos\frac{k_1}{2} & 0 & \mathrm{e}^{-\mathrm{i}\phi}\cos\frac{k_2}{2} \\
\mathrm{e}^{-\mathrm{i}\phi}\cos\frac{k_1+k_2}{2} & \mathrm{e}^{\mathrm{i}\phi}\cos\frac{k_2}{2} & 0
\end{pmatrix}.
$$
It is obvious that this Hamiltonian is invariant under the rotation for $\pi$. A rotation for $2\pi/6$ and reordering basis vectors [$(RGB) \rightarrow (GBR)$] leaves the Hamiltonian unchanged. This means that the Berry curvature is invariant under rotation for $2\pi/6$. By adding an arbitrary vector of reciprocal lattice to $\mathbf{k}$, off-diagonal elements might change their sign which is of no influence on the Berry curvature.

\begin{figure}[hbt]\centering
\includegraphics[width=0.75\linewidth]{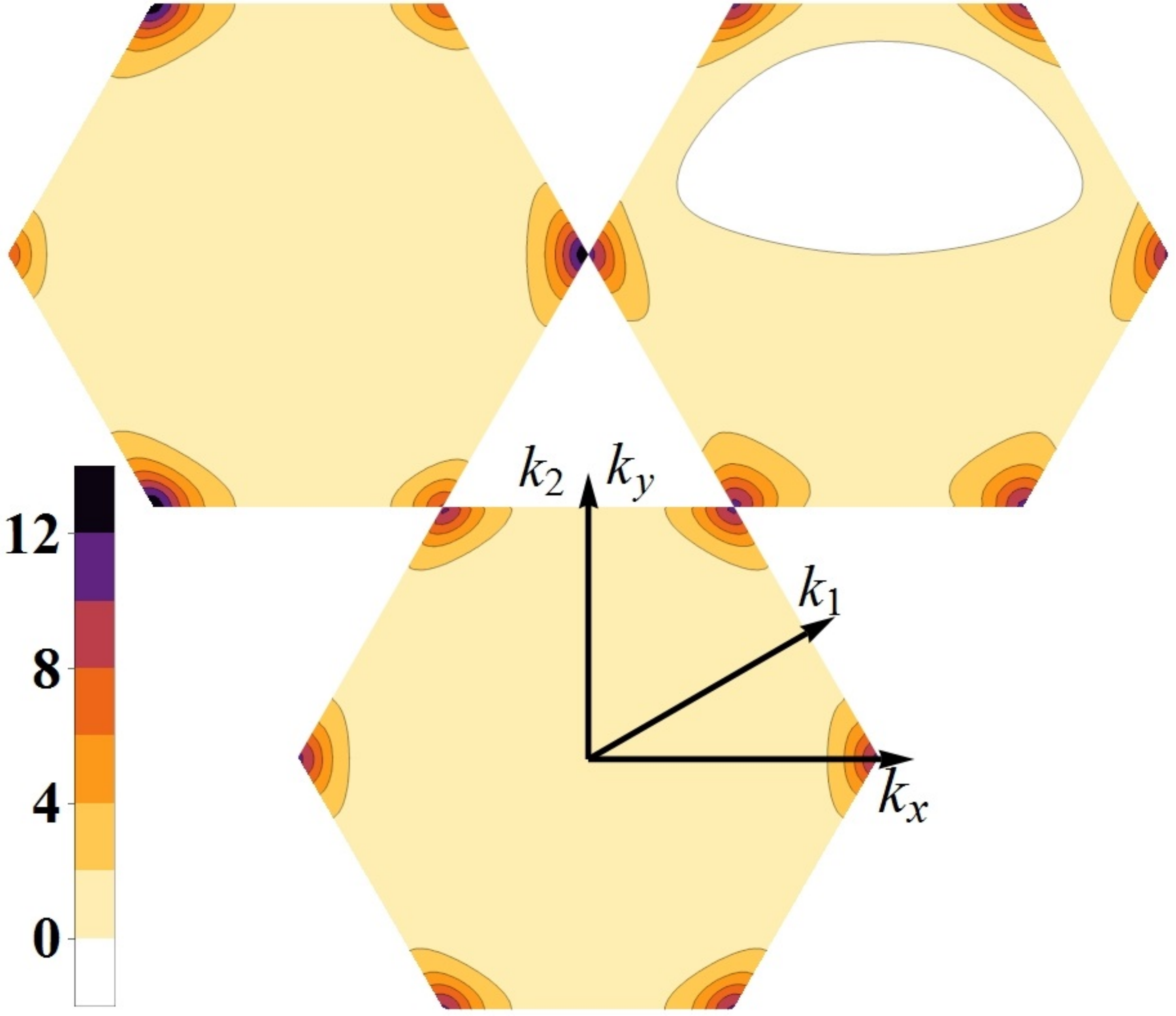}
\caption{\label{FKBerry} (Color online) Berry curvatures of the lowest energy band for the kagome model (parameters are $t_1=1$ and $\phi=\pi/4$) in the Brillouin zone. Up-left is of the Bloch Hamiltonian for the light-grey (see Fig.~\ref{Fkagome}) and up-right for the dark-grey unit cell. Down is for the generalized Bloch Hamiltonian constructed here.}
\end{figure}
The Berry curvature of the lowest energy state of the generalized Bloch Hamiltonian is given in Fig.~\ref{FKBerry} (down). Notice that it has the full symmetry of the crystal. Above are the Berry curvatures, also for the lowest energy state, of the Bloch Hamiltonians for light (left) and dark-grey (right) unit cells (see Fig.~\ref{Fkagome}). Again, the average of any of these two Berry curvatures and ones rotated for $\pi$ (i.e. of the Bloch Hamiltonians for unit cells rotated for $\pi$, which is the symmetry of the model), is the Berry curvature of the generalized Bloch Hamiltonian. Notice that the Berry curvature for the light-grey unit cell is symmetric under rotation for $2\pi/3$, like that unit cell, while for the dark-grey unit cell has only the translational symmetry.

\begin{figure}[hbt]\centering
\includegraphics[width=0.7\linewidth]{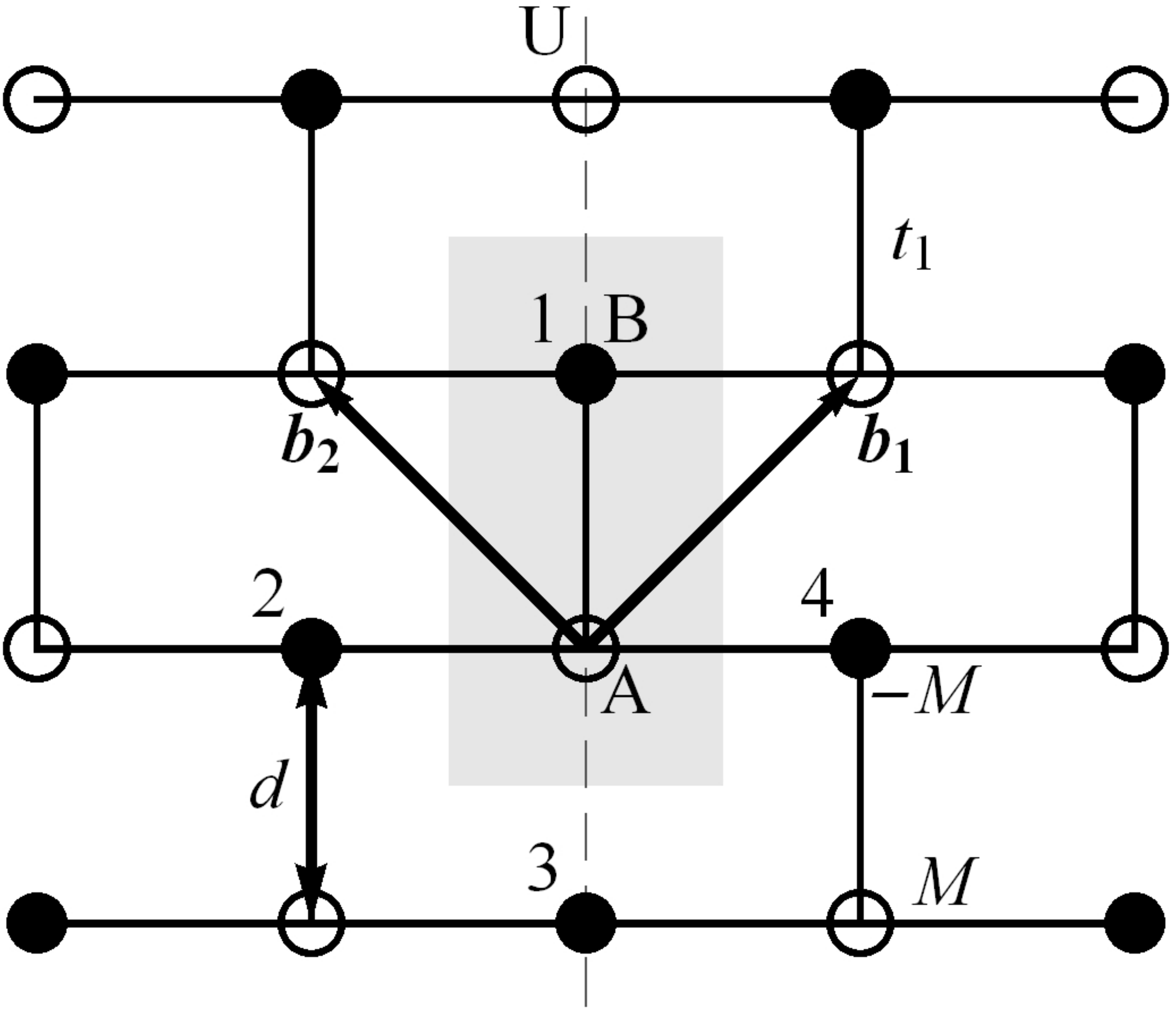}
\caption{\label{Fbrickwall} The brick-wall lattice model ($t_2=0$) with two sublattices $A$ (circles) and $B$ (disks), with on-site energies $M$ and $-M$, respectively. The nearest neighbors hoping amplitude is $t_1$, only between connected atoms. The distance of NNs is $d$. Dashed line is the group element $U$, horizontal axis of rotation for $\pi$. Vectors $\mathbf{b}_1$ and $\mathbf{b}_2$ are direct lattice vectors. Grey area is one choice for unit cell. Other choices are $A$ atom from that unit cell with one of the nearest neighbors, $B$ atom 2, 3, or 4.}
\end{figure}

Finally, the third example is the brick-wall model~\cite{Dmitry-PhysRevLett.110.165304}. In order to illustrate the case with anisotropic interactions with non-zero vectors $\mathbf{a}_{j}^0$, we took model without next nearest neighbors hoping amplitude $t_2=0$ (see Fig.~\ref{Fbrickwall}). This model is invariant under the 2D translations along $\mathbf{b}_1=d\sqrt{2}(1,0)$ and $\mathbf{b}_2=d\sqrt{2}(0,1)$, where $d$ is the NN distance, and the rotation for $\pi$ about horizontal $U$ axis (centerline of $\mathbf{b}_1$ and $\mathbf{b}_2$).
The general irreducible representation for these generators is
\begin{align}
D(U) &= \begin{pmatrix} 0&1\\1&0\end{pmatrix},\nonumber\cr
D(\mathbf{b}_1) &= \mathrm{diag}(\mathrm{e}^{-\mathrm{i}k_1}, \mathrm{e}^{-\mathrm{i}k_2}),\nonumber\cr
D(\mathbf{b}_2) &= \mathrm{diag}(\mathrm{e}^{-\mathrm{i}k_2}, \mathrm{e}^{-\mathrm{i}k_1}).\nonumber
\end{align}
The stabilizers for the atoms $A$ and $B$ in the grey UC are equal $S_A=S_B=\{ e , U \}$. Vectors from the range of the projector for this stabilizer are
$$
|A\rangle\rangle = \mathrm{e}^{-\mathrm{i}\mathbf{k}\cdot\mathbf{a}_{A}^0} \begin{pmatrix} 1 \\ 1 \end{pmatrix},\,
|B\rangle\rangle = \mathrm{e}^{-\mathrm{i}\mathbf{k}\cdot\mathbf{a}_{B}^0} \begin{pmatrix} \mathrm{e}^{\mathrm{i}(k_1+k_2)/2} \\ \mathrm{e}^{\mathrm{i}(k_1+k_2)/2} \end{pmatrix},
$$
with an arbitrary $\mathbf{a}_{A/B}^0$ along $\mathbf{b}_1+\mathbf{b}_2$, $\mathbf{a}_{A} = \mathbf{0}$, and $\mathbf{a}_{B} = (\mathbf{b}_1+\mathbf{b}_2)/2$. Since one of the vectors $\mathbf{a}_{A/B}^0$ is redundant, we can take $\mathbf{a}_A^0 = \mathbf{0}$ and let $\mathbf{a}_B^0 = c(\mathbf{b}_1+\mathbf{b}_2)$ with an arbitrary constant $c$. With the vectors $|A/B\rangle\rangle$, the generalized Bloch Hamiltonian matrix elements are $H^A_A = M = - H^B_B$ and
$$
{H_A^B}^* = H_B^A = \mathrm{e}^{-\mathrm{i}\mathbf{k}\cdot\mathbf{a}_{B}^0} t_1 ( 2 \cos\frac{k_1-k_2}{2} + \mathrm{e}^{\mathrm{i}\frac{k_1+k_2}{2}} ).
$$
Notice that the Hamiltonian matrix elements are invariant under the rotation for $\pi$ about $U$, while adding general reciprocal lattice vector $\mathbf{G} = m_1\mathbf{g}_1 + m_2\mathbf{g}_2$ to $\mathbf{k}$, brings a constant phase $\pi + c (m_1+m_2)2\pi$ into off-diagonal elements.

Different vectors $\mathbf{a}_{B}^0$ provide different gauges. If $\mathbf{a}_B^0$ is equal to zero we have for the generalized Bloch Hamiltonian the Hamiltonian in periodic gauge. If we choose $\mathbf{a}_B^0$ to be equal to the position of the $B$ atoms 1 or 3 [$\mathbf{a}_B^0 = \pm(\mathbf{b}_1+\mathbf{b}_2)/2$, with plus for the atom 1 and minus for 3], then the generalized Bloch Hamiltonian is the Bloch Hamiltonian for the unit cells with central atom $A$ and $B$ atoms 1 or 3, respectively. Notice that the symmetry does not allow a construction of the Bloch Hamiltonian for the unit cells with the $B$ atoms 2 or 4.

\begin{figure}[hbt]\centering
\includegraphics[width=0.75\linewidth]{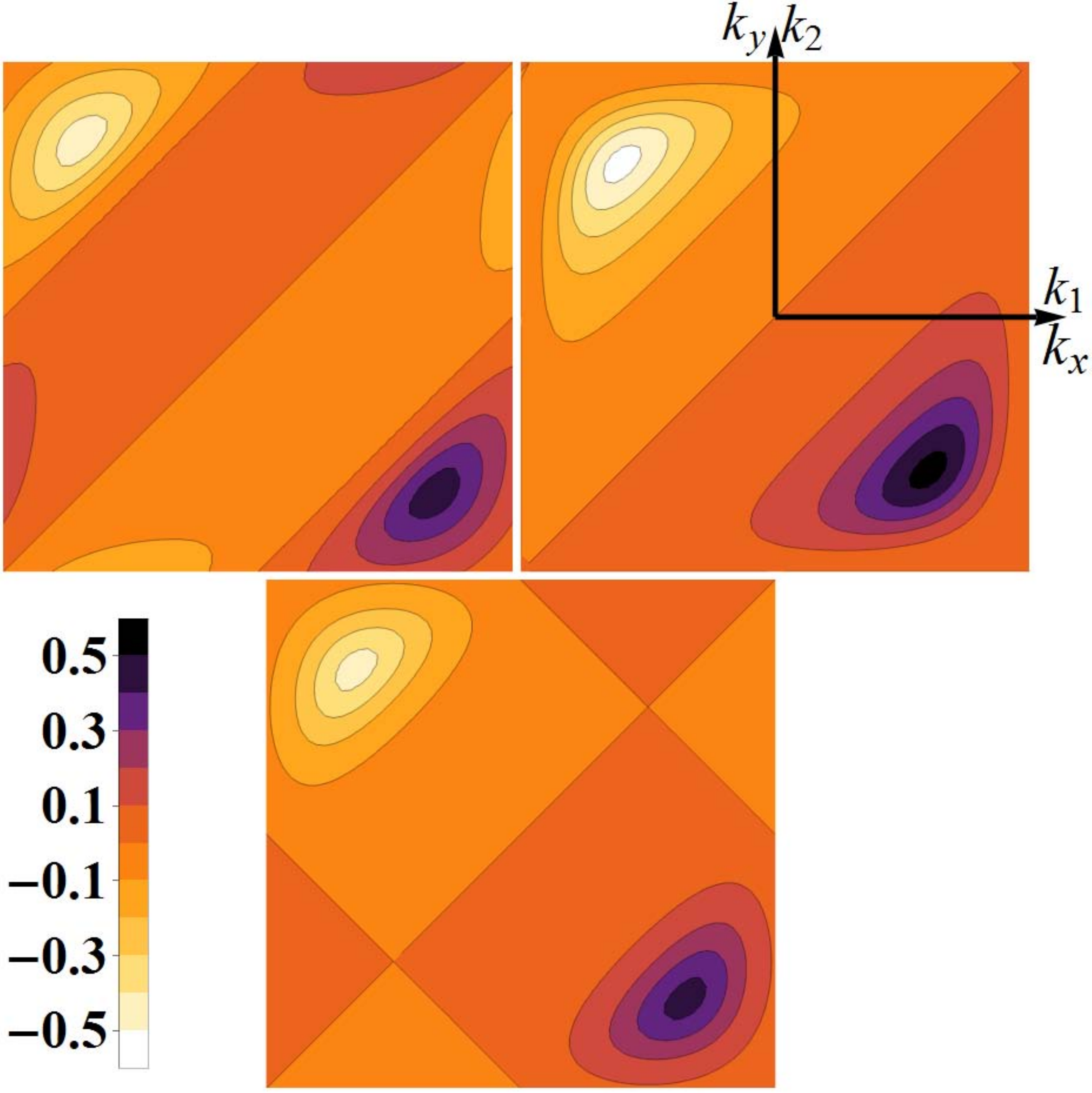}
\caption{\label{FBBerry} (Color online) Berry curvatures of the lowest energy band for the brick-wall model (parameters are $M=1$, $t_1=1$, and $t_2=0$) in the Brillouin zone. All the curvatures are for the unit cells containing atom A in the origin (see Fig.~\ref{Fbrickwall}). Up-left is for the Bloch Hamiltonian for the grey unit cell [$\mathbf{a}_B^0 = (\mathbf{b}_1 + \mathbf{b}_2)/2$] and up-right for the unit cell with $B$ atom 3 [$\mathbf{a}_B^0 = -(\mathbf{b}_1 + \mathbf{b}_2)/2$]. Down is for the generalized Bloch Hamiltonian with $\mathbf{a}_B^0=\mathbf{0}$, i.e. the Hamiltonian in the periodic gauge.}
\end{figure}

Again, all the possible (by varying $\mathbf{a}_B^0$) Berry curvatures for the generalized Bloch Hamiltonian have the symmetry of the model. The Berry curvature of the lowest energy state of the Hamiltonian in periodic gauge ($\mathbf{a}_B^0 = \mathbf{0}$) is given in the lower part of Fig.~\ref{FBBerry}.
Others are the curvatures for the lowest energy state of the Bloch Hamiltonians constructed with different unit cells; atom $A$ being one in the origin and the $B$ atom being 1 or 3. The unit cells with atom 1 and 3 have $U$ symmetry and therefore their Berry curvatures, up-left and up-right in Fig.~\ref{FBBerry}, respectively, have the following transformation property $U B(\mathbf{k}_1,\mathbf{k}_2)=-B(\mathbf{k}_2,\mathbf{k}_1)$. Although the symmetry of the curvatures is the same, the average of these two curvatures is the curvature of the generalized Bloch Hamiltonian for $\mathbf{a}_B^0 = \mathbf{0}$, i.e. the Hamiltonian in the periodic gauge. As a consequence, here, like in the Haldane model, the Berry curvature of any eigenstate of the Hamiltonian in the periodic gauge is average of the curvatures of the Bloch Hamiltonians constructed for the unit cells made of one fixed atom and its  nearest neighbors.

\section{Summary}
The Bloch theorem is generalized to the case when a crystal is invariant under more symmetry operations than just translations. The method applied can be used even in the case of the systems without translational symmetry like molecules. In this generalization the method provides the most symmetric, in the reciprocal space, Hamiltonian and Berry curvature. The resulting Hamiltonian coincides with the Hamiltonian defined by the periodic gauge. In addition it is shown, for any energy band, the curvature of the Hamiltonian in the periodic gauge is equal to the average of the curvatures of the Bloch Hamiltonians for all possible unit cells.

\section*{Acknowledgments}
We would like to thank B. Nikoli{\'{c}} for many useful comments and suggestions. This work was supported by the Ministry of Education, Science, and Technological Development of the Republic of Serbia under projects No. ON171027, ON171031, and ON171017.

\bibliography{gBt}

\end{document}